\documentclass[aps,prx,reprint,superscriptaddress, longbibliography]{revtex4-1}
\usepackage{H1}
\usepackage{amsmath}
\usepackage[pdftex,colorlinks=true]{hyperref}
\hypersetup{
citecolor = blue
}
\usepackage[normalem]{ulem}
\usepackage{amssymb}
\usepackage{amsthm}
\usepackage{amsfonts}
\usepackage{bbm}
\usepackage{array}

\begin{document}

\title{Local constraints can globally shatter Hilbert space: a  new route to quantum information protection}
\author{Vedika Khemani}
\affiliation{Department of Physics, Harvard University, Cambridge, MA 02138, USA}
\author{Rahul M. Nandkishore}
\affiliation{Department of Physics and Center for Theory of Quantum Matter, University of Colorado, Boulder, CO 80309}

\begin{abstract} 
We show how local constraints can globally `shatter' Hilbert space into subsectors, leading to an unexpected dynamics with features reminiscent of both many body localization and quantum scars. A crisp example of this phenomenon is provided by a `fractonic circuit' - a model of quantum circuit dynamics in one dimension constrained to conserve both charge and dipole moment. We show how the Hilbert space of the fractonic circuit dynamically fractures into disconnected \emph{emergent} subsectors \emph{within} a particular charge and dipole symmetry sector. A large number of the emergent subsectors, exponentially many in the size of the system, have dimension one and exhibit strictly localized quantum dynamics---even in the absence of spatial disorder and in the presence of temporal noise. Exponentially large localized subspaces can be proven to exist for any one dimensional fractonic circuit with finite spatial range, and provide a potentially new route for the robust storage of quantum information. Other emergent subsectors display non-trivial dynamics and may be constructed by embedding finite sized non-trivial blocks into the localized subspace. The shattering of a particular symmetry sector into a distribution of dynamical subsectors with varying sizes leads to the coexistence of high and low entanglement states, \emph{i.e.} this provides a general mechanism for the production of quantum many body scars. We discuss the detailed pattern of fracturing and its implications. We also discuss other mechanisms for similarly shattering Hilbert space. 
\end{abstract}
\maketitle

\normalsize

\section{Introduction}

This paper lies at the intersection of two interesting streams of contemporary research: the study of quantum information and its protection, and the study of nonequilibrium dynamics in quantum systems. A major theme of interest in the former line of work is the quest for new models or phases of matter that protect quantum information (see \cite{Loss} and references contained therein). While examples of \emph{thermally stable} self correcting quantum memories exist in four or more dimensions, such as the toric code~\cite{Kitaev}, the effects of finite temperature are more debilitating in lower dimensions.  For example, while topological phases in two dimensions are known to possess a ``logical subspace" into which information may be encoded in a manner immune to local perturbations, this protection is only afforded at zero temperature~\cite{Kitaev}. The recent discovery of `fracton' phases of quantum matter \cite{chamon, haah, fracton1, fracton2, sub, genem, fractonarcmp} has opened a new direction in this quest.  Fracton phases in three spatial dimensions also have a protected logical subspace (immune to local perturbations at zero temperature \cite{kimhaah}), but of a size which grows exponentially in the linear extent of the system---unlike conventional topological phases, where the logical subspace has constant dimension. Nevertheless, it is believed that three dimensional `fractonic' phases do thermalize at non-zero temperatures, such that they do not constitute ideal quantum memories at finite temperature \cite{SivaYoshida,Prem}. 

A parallel major theme of investigation involves the study of {\it thermalization} in isolated quantum many body systems. The cornerstone of this understanding is the Eigenstate Thermalization Hypothesis (ETH)\cite{Deutsch,Rigol,Srednicki}. The strong form of the ETH holds that \emph{all}  many-body energy eigenstates are thermal (the weak form of the ETH, in which only almost all eigenstates are thermal, is known to not be sufficient to guarantee thermalization \cite{Kollath}).  In recent years,  interest has grown in systems that violate the ETH. One well known class of `counterexamples' to ETH is provided by integrable systems, which possess an extensive number of conserved quantities, and thermalize instead to a generalized Gibbs ensemble \cite{Vidmar}. The other well studied counter-example is many-body localization (MBL)\cite{GMP,BAA,mblarcmp,EhudMBLReview}, driven by disorder, in which case essentially all eigenstates are nonthermal, characterized by an extensive number of {\it emergent} local integrals of motion~\cite{Huse14, Serbyn13cons}. As a result, MBL systems never reach thermal equilibrium and retain local memory of their initial conditions for arbitrarily late times --- a feature that can preserve quantum information even at finite energy-densities, and could potentially have use in developing new technologies such as quantum memories. 

Still more recently, a new type of counterexample to strong ETH has been observed, which now goes under the name of ``quantum many-body scars" \cite{AKLT1, ShiraishiMori, turner2018weak, ck, Serbyn2, LinMotrunich, moudgalya2018entanglement, TDVPScars, ChoiAbanin, ok2019topological,  ck, Misha51atom, lerose, ichinose, konik, Iadecola,PollmannDimerScars, LeroseConfinement, KonikConfinement, MartinScars,TurnerRevivals}. Quantum many-body scars are  loosely defined as a small number of non-thermal eigenstates (measure zero in the thermodynamic limit) embedded into an otherwise thermal spectrum. The presence of these ``scar" states can lead to distinct signatures in quench experiments if the initial states have high overlap with the scar states, as was recently observed in an experiment on Rydberg atoms~\cite{Misha51atom}. While it is known how scars may be embedded, by hand, into a thermal many-body spectrum in certain special classes of Hamiltonians \cite{ShiraishiMori,ShiraishiAKLT}, there is still little understanding of the general principles that could give rise to scars~\cite{TDVPScars, ck,lerose, ichinose, konik, Iadecola, LeroseConfinement, KonikConfinement} , whether the phenomenon survives weak perturbations~\cite{ck}, and whether and when scars should be expected to arise in generic Hamiltonians in the thermodynamic limit, making this a very active area of research. 

While the origin of scars, in general, is still largely unexplained (including in the Rydberg model), several known models with ``exact" or ``perfect" scars (in particular, those in \cite{ShiraishiMori, ChoiAbanin, ok2019topological}) can be understood as examples of the construction in Ref~\cite{ShiraishiMori}. A useful perspective on Ref~\cite{ShiraishiMori}, connecting it to this present work, is to note that the construction therein leads to scars via a dynamical \emph{fracturing} of the Hilbert space into disconnected emergent subsectors, even in the absence of explicit conservation laws.  If one or a few of these emergent subspaces can be spanned using only a small number of low-entanglement basis states (that constitute a vanishing fraction of the full Hilbert space), then eigenstates living in these subspaces necessarily have low entanglement. These will violate strong ETH when they coexist at the same energy densities as the other, thermal, eigenstates.  However, barring the relatively fine-tuned setup in~\cite{ShiraishiMori}, general conditions that lead to emergent Hilbert space fracture are not known.  Our work adds to this budding literature by furnishing a robust class of constrained models where such Hilbert space fracture, and hence scarring, can be proven to exist on very general grounds --- thus providing at least one concrete (and robust) mechanism for obtaining scars.

More generally, in this manuscript, we marry together two lines of research by demonstrating how certain local constraints can give rise to a dramatic fracturing of Hilbert space into {\it exponentially} many emergent dynamical subsectors --- whence the word ``shatter".   Our models are constructed  in { one} spatial dimension and, unlike two dimensional topological phases or three dimensional fracton phases, the localized subspace is not limited to zero temperature and is robust to both temporal noise and local perturbations that respect the constraints. 
This raises the tantalizing possibility that quantum information may be robustly encoded in these subpaces without any loss.
Unlike conventional MBL, the `localization' here involves only a measure zero fraction of the full spectrum, and requires neither disorder nor energy conservation. 

The cleanest setting for illustrating this phenomenon is the `fractonic random circuit' introduced in \cite{pai2018localization}. This is a model of quantum circuit dynamics \cite{Nahum1, Nahum2,  Keyserlingk2, KhemaniVishHuse,  Keyserlingk1} constrained to conserve both a $U(1)$ charge and its dipole moment. We examine the pure state dynamics in this system and find that the
Hilbert space in this model ``shatters" into an exponentially large number of dynamical subsectors \emph{within} a given charge and dipole symmetry sector, giving rise to a breakdown of ergodicity and a violation of ETH as conventionally understood. 
We identify, in particular, an \emph{exactly} localized subspace (exponentially large in system size) which can be labeled by state-dependent emergent local integrals of motion, and into which quantum information may be robustly encoded. 

Indeed, one of our main results is an analytic \emph{proof} that the conservation of charge and dipole moment, along with spatial locality, is sufficient to produce exponentially many strictly inert states which live in subspaces of dimension exactly exactly equal to one. These looks like simple product states in the computational basis, and have zero entanglement. 
The existence of these states is particularly is particularly striking because the violation of ETH typically requires $O(L)$ local, mutually commuting conservation laws, while our systems have only two. We also identify how to systematically construct additional dynamical sectors with non-trivial dynamics by embedding non-trivial blocks into the localized background. The upshot is the co-existence, in a single symmetry sector with a particular charge and dipole moment, of dynamical sectors of various sizes and hence of both high and low entanglement states, i.e. the shattering of Hilbert space produces quantum scarring. We comment on some of the implications for dynamics.   


Because our results on fracture only requires charge and dipole conservation and spatial locality, the results are stable to all perturbations obeying these constraints. This is evidenced in our chosen circuit models by the fact that the unitary gates generating the dynamics may be chosen randomly with respect to Haar measure and are not fine tuned in any way. In particular, our results no not require disorder in space, are are stable even to temporal fluctuations which typically kill MBL. 

We note that physics analogous to fracture has also been observed in other models. These include, for example, the Fermi-Hubbard model and its cousins~\cite{BernevigFH, IadecolaFH},  models
 with kinetic constraints (including in classical settings)~ \cite{LanGarrahan, OlmosLesanovsky, Gopalakrishnan_automata},  and dimer models~\cite{dimerfracture}. However, although constraints can lead to disconnected subsectors of Hilbert space in these cases, there is generally no principled way to examine the stability of fracturing in these models to the addition of perturbations or noise. For example, the simplest kinetically constrained models comprise spin 1/2 systems in which the spin on a site can flip if certain conditions are obeyed by its neighbors, for instance if both neighbors are down. However, there is no unique way to ``extend" such models, for example, to include the effect of further neighbor spins. Likewise, the dynamics in quantum dimer models come from certain ``flippable" plaquettes which are lattice dependent~\cite{QDM_review}. While allowing for longer flippable loops decreases fracture~\cite{dimerfracture}, there are no general results on how the number of disconnected sectors scales with such perturbations. By contrast, scars in our model comes from a clear physical origin --- the conservation of charge and dipole moment --- which furnishes a natural class of symmetry respecting perturbations. 
 We thus expect our results on fractonic circuits, including classical stochastic versions thereof, to also be of interest to future studies on kinetically constrained models.

Finally, while the `fractonic' random circuit provides a clean example of local constraints shattering Hilbert space, we also point out that not all local constraints act in this way. Nevertheless, the `shattering' of Hilbert space is not particular to fractonic circuits, and we also discuss some other (not obviously fractonic) examples of circuits which also exhibit a shattered Hilbert space. We conclude with a discussion of some open directions. 

\section{The model}
We work with the model of quantum circuit dynamics introduced in \cite{pai2018localization}. The Hilbert space consists of a one dimensional chain of $S=1$ quantum spins of length $L$, acted upon by local unitary gates which locally conserve both charge ($Q = \sum_j S_j^z$) and dipole moment ($P = \sum_{ j} j S^z_j,$), where $j$ is a site label. We can work with basis states in the $S^z$ basis, as these are eigenstates of both $P$ and $Q$.  On each site, the allowed values of $S^z$ are $|+\rangle, |-\rangle, |0\rangle$. The twin conservation laws greatly restrict the allowed movement of charges (fractons), and this restricted movement is a defining property of fracton phases. For example, a single $+$ or $-$ charge on site $r$ has dipole moment $P= \pm r$. Such a charge cannot simply ``hop" to the left or right, because such a movement changes the net dipole moment by one unit. On the other hand, bound states of charges  or ``dipoles" of the form $(-+)$ have net charge zero and net dipole moment $P=\pm1$ indepedent of position, and these can move freely through the chain. Additionally, dipoles can enable the movement of charges, because a charge can move if it simultaneously emits a dipole to keep $P$ unchanged: $|0+0\rangle \rightarrow |+-+\rangle$.

\begin{figure}[t]
\centering
 \includegraphics[width=\columnwidth]{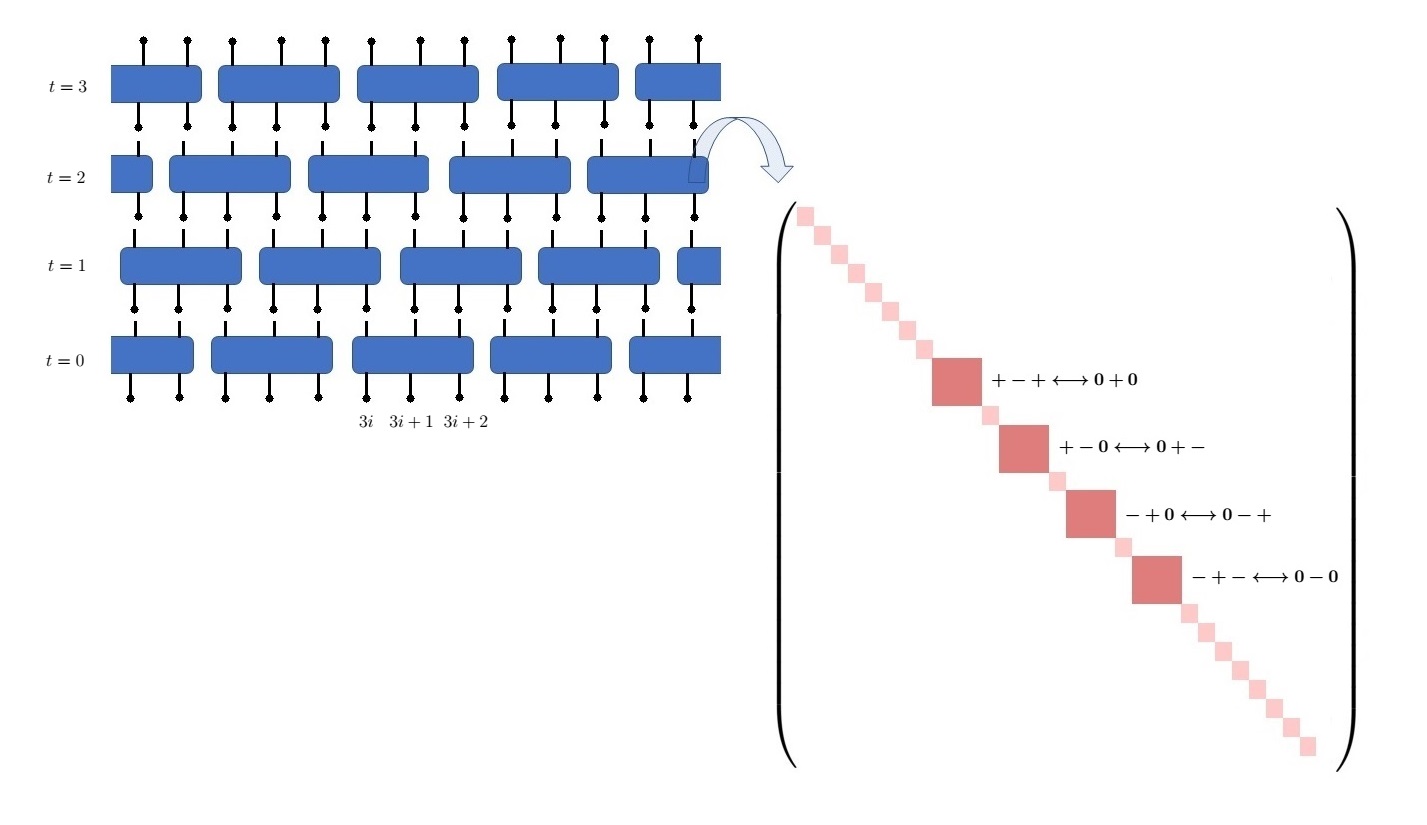}
 \caption{Fractonic random unitary circuit: each site is a three-state qudit. Each gate (blue box) locally conserves charge $Q = \sum_j S^z_j$ and dipole moment $P = \sum_j j S^z_j$ of the three qudits it acts upon. The block diagonal Haar-random unitary with its nontrivial blocks is also shown. Figure taken from \cite{pai2018localization}.}
 \label{fig:floquetCircuit}
 \end{figure}

The simplest realization of these rules is provided by circuits with three site unitary gates, which take the form of $27\times27$ matrices as shown in Fig.\ref{fig:floquetCircuit}. The charge and dipole moment conservation lead to a block diagonal structure in the gates. Notably, there are only four non-trivial two by two `blocks,' each of which is a random unitary drawn independently from the Haar measure on $U(2)$, while the rest of the matrix is diagonal (pure $U(1)$ phase). We will begin our analysis with a discussion of this simple circuit with three site gates, but we will prove that the key results are robust for any finite gate size (while also flagging some special features that do depend on gate size). 
We note that while \cite{pai2018localization} considered a circuit that was random in both space and time, this is not important for our purposes - our results hold just as well if the circuit is uniform in space (translation invariant), and/or if it is periodically repeated in time (Floquet). In all that follows, we work with a circuit that is translation invariant, since this makes our central result localization yet more dramatic. We also work with a circuit that is stroboscopically repeated with period 3, since this allows us to meaningfully discuss eigenstates. However, we emphasize that our basic results require neither translation invariance in space, nor periodicity in time. 

 This circuit has only two symmetries: charge conservation and dipole moment, and the `symmetry sectors' of the theory are correspondingly labelled by just two quantum numbers: charge $Q$ and dipole $P$. In the Floquet version of the model, three staggered ``layers'' of the circuit are chosen independently, but the layers are then repeated in time. The time evolution operator for one Floquet period is given by 
$U^F  = U_3 U_2 U_1$, where 
\begin{equation}
  U_n=\begin{cases}
    \prod_{i}U^n_{3i,3i+1,3i+2} & \text{if $n = 0$}\\
    \prod_{i}U^n_{3i-1,3i,3i+1} & \text{if $n = 1$}\\
    \prod_{i}U^n_{3i-2,3i-1,3i} & \text{if $n= 2$},
  \end{cases}
\end{equation}
where the gates $U^{1}, U^{2}$ and $U^{3}$ are chosen at random for a given realization, but remain fixed throughout the run corresponding to that realization.  We work throughout with open boundary conditions. In certain layers of the circuit, there may be sites near the boundary that are acted on trivially (pure phase) but the Floquet operator as a whole acts non-trivially on every site. 

\begin{figure}
 \centering
\includegraphics[width=\columnwidth]{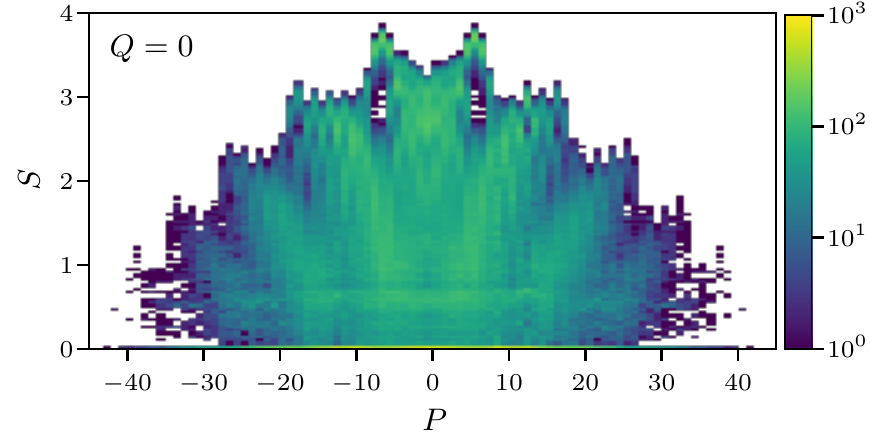}
\caption{ Entanglement entropy of eigenstates as a function of dipole moment for a system with $L=13$ sites, in the symmetry sector with total charge $Q=0$. The color-bar denotes the number of eigenstates with entanglement entropy $S$ and a given $(Q,P)$. For each symmetry sector $(Q,P)$, there is a co-existence of low and high entanglement eigenstates, in sharp contrast to the usual behavior expected from the eigenstate thermalization hypothesis.\label{EE} }
 \end{figure}

Before presenting our analytic proofs,  we  illustrate the unusual properties of this model by numerically studying the eigenstates of the Floquet unitary within each symmetry sector. For each eigenstate $|\psi \rangle$ we construct a density matrix $\rho = |\psi\rangle \langle \psi|$, and extract the half-chain entanglement entropy $S$ according to $S = - \mathrm{Tr}_B \rho \log \rho$, where the trace is over half the chain. In Fig.~\ref{EE} we plot the entanglement entropy of the eigenstates for a system of size $L=13$, in total charge $Q=0$ sector, as a function of dipole moment $P$. We note that the states with maximal charge have $Q=\pm L$, so $Q=0$ corresponds to the middle of the many body spectrum, where we could expect the ETH to apply in a translation invariant and not conventionally integrable model. However, in every symmetry sector $(Q,P)$ we find a combination of low and high entanglement eigenstates, in sharp contrast to the usual expectations from eigenstate thermalization, but analogous to the phenomenon of quantum many body scars. As we will show, this apparent violation of the ETH arises from the shattering of Hilbert space. 
 
\section{Shattering of Hilbert space}
We now demonstrate how the local constraints fracture Hilbert space, giving rise to an exponentially large number of emergent dynamical subsectors. By contrast, note that the twin conservation laws of charge and dipole moment only lead to $O(L^3)$ explicit symmetry sectors, labeled by the values of charge and dipole moment ranging from $Q = \{-L,  \cdots, L\}$ and $P = \{-\frac{L(L-1)}{2}, \cdots \frac{L(L-1)}{2}\}. $

\subsection{Localized eigenstates}
\label{inert}
In this section, we show how all local fractonic circuits have exponentially many \emph{exactly} localized inert states, labeled by state dependent local integrals of motion (despite the absence of spatial randomness). These constitute emergent subsectors of dimension exactly one. Noteably, these inert states are product states of charge ($i.e.$ product states of $S^z$), so these are exceptionally simple, physically realizable states. These states are eigenstates of the Floquet fractonic circuit with zero entanglement, while they are left invariant by circuits that are random (\emph{i.e.} non-repeating) in time, thereby also demonstrating robustness to temporal noise. 

We start with an analytic proof which shows that the combination of $Q,P$ symmetries together with locality is enough to give exponentially many strictly inert states. The construction in our proof is extremely physical, and furnishes a strict lower bound on the number of inert states. We then provide an inductive, though less physical, method which allows us to count the actual number of inert states. 

 \begin{figure}
 \centering
\includegraphics[width=\columnwidth]{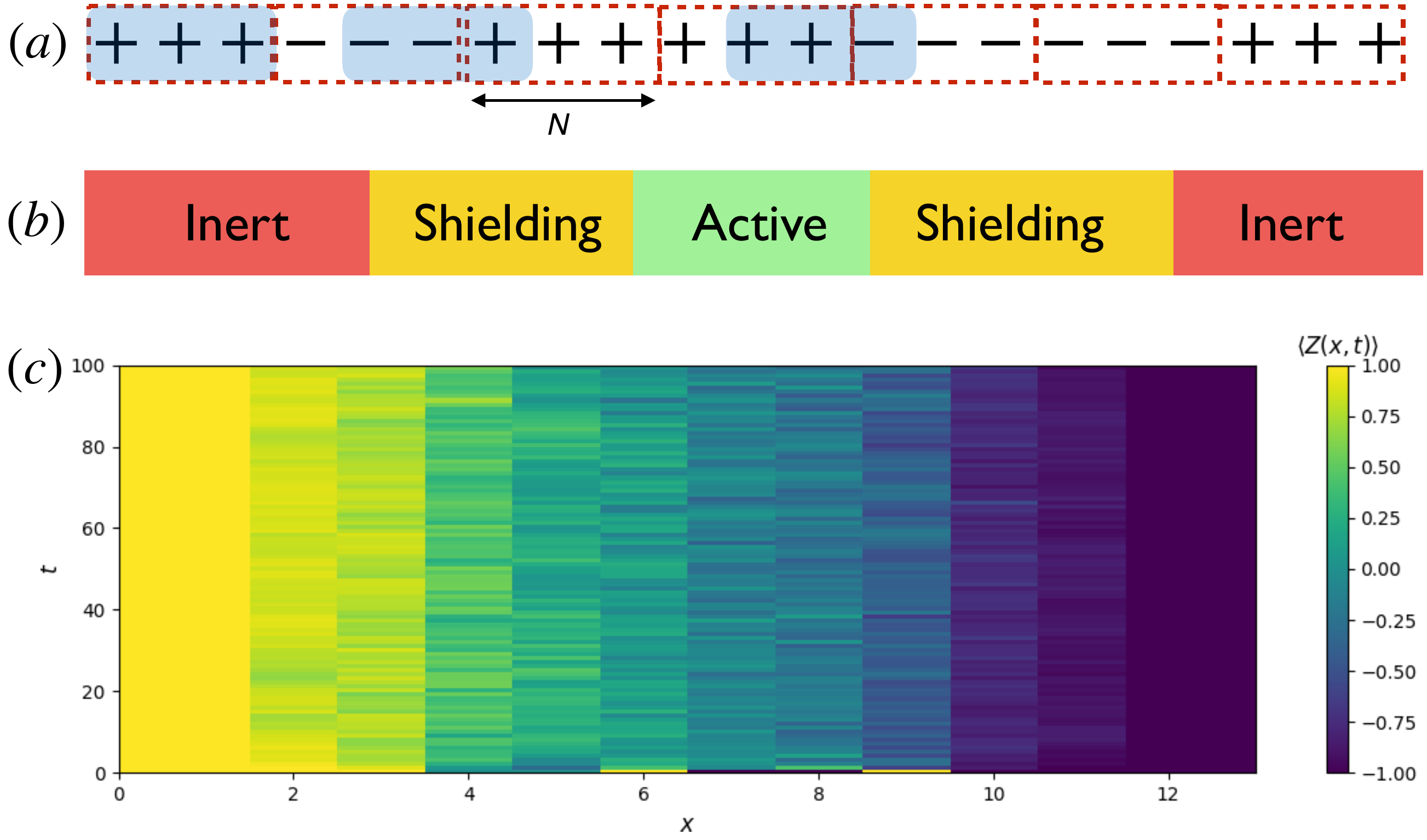}
\caption{ \label{fig:inert} (a) Exponentially many strictly inert states in a model with range $N$ can be constructing by dividing the system into size $N$ blocks, and randomly picking each block to be of extremal positive or negative charge. A range $N$ gate (blue rectangles) acting on such a state locally sees either a configuration of maximal charge, or a configuration of maximal dipole moment for a given charge --- and hence is forbidden from making any local rearrangements. (b) Dynamical subspaces of varying sizes can be constructed by embedding ``active", \emph{i.e.} non-inert, blocks into inert backgrounds. As long as the active block has a finite size, it can be prevented from melting the inert regions by surrounding it with ``shielding" regions of equal or greater size. (c) Dynamics of charge $\langle S^z_x(t)\rangle$ starting from an initial state with a central active region surrounded by shielding regions. We see that the central region thermalizes, but isn't able to melt the boundary spins which remain inert.  }
 \end{figure}

Consider a model with charge and dipole symmetries, and finite range (gate-size) $N$. Now, note that {\it any} pattern that interconverts between locally `all plus' and locally `all minus,' with domain walls between `all plus' and `all minus' regions at least $N$ sites apart, must be inert. These are states of the form $|++++++-----++ \cdots\rangle$ (\emph{cf.} Fig~\ref{fig:inert} (a)). This follows because every gate acting on such a state straddles either zero or one domain walls. If it straddles zero domain walls, then it acts locally on a block with extremal charge, which is obviously inert. If it straddles one domain wall, then it acts on a block with extremal dipole moment {\it given its charge}, and this must also be inert. The inertness of the latter kind of block follows because it is made up of only $+$ and $-$ charged sites, and the only charge conserving moves that one can make are (i) to reshuffle $+$ and $-$ charges and (ii) to delete $+$ and $-$ charges in pairs and replace them by zeros. However, if {\it every} $+$ charge is to the right of {\it any} $-$ charge (or vice versa) then any such move necessarily changes the dipole moment, and so is forbidden. 

One can then straightforwardly lower bound the size of the exactly localized subspace for circuits with gate-size $N $ by dividing the system up into blocks of length $N$, and allowing each block to be either `all plus' or `all minus.' This yields an inert subspace of dimension at least $2^{L/N} = c^L$, where $c = 2^{1/N}$. This is exponentially large in system size for {\it any} finite gate size $N$, and cleanly illustrates how simultaneously conserving charge and dipole moment provably leads to the emergence of exponentially large localized subspaces into which information may be robustly encoded.  

Note that the bound above is not tight; for $N=3$ it predicts a localized subspace of dimension at least $1.25^L$, whereas a more careful counting, done below, gives a localized subspace of dimension $2.2^L$. Nevertheless, it is sufficient to establish the existence of an exponentially large, robust, localized subspace for any finite gate size. Each of the inert states in this subspace can be labeled by \emph{state-dependent} local integrals of motion corresponding to the local values of charge and dipole moment. Also, note that this type of localization does not require disorder - indeed it occurs even in a circuit that is translationally invariant in the thermodynamic limit and survives temporal noise, as long as the constraints are obeyed.

We now turn to a more precise counting of the inert states. Exactly localized eigenstates may be constructed in the thermodynamic limit using an inductive method. We demonstrate this for the circuit with range three unitary gates. For system size $L=3$, there is only one gate acting, and there are exactly $19$ product states (in the charge basis) which have trivial dynamics, and are hence localized - these are the $19$ states acted upon by trivial blocks of the constrained random unitary in Fig.\ref{fig:floquetCircuit} (e.g. the state $|00+\rangle$). These states do not mix with the rest of the Hilbert space, and are hence `inert,' lying in a subsector with dimension one. Meanwhile, if a state is inert  in a system of size $L$, then it will remain inert when an additional degree of freedom is added if the final two degrees of freedom of the $L$ site system and the additional degree of freedom collectively form one of the `inert' configurations of an $L=3$ site system. This is because the only ``new" dynamics in the presence of the additional spin comes from the addition of a single three site unitary gate acting on the three spins formed by the added spin and the two penultimate spins of the length $L$ chain. Importantly, for {\it any} inert state of an $L$ site system, there is at least one choice of spin state for the added spin (and sometimes more than one), which leaves the resulting state in the $L+1$ site system also inert. Specifically, an inert state in a system of size $L$ remains inert upon addition of another degree of freedom if the conditions tabulated in Table \ref{threesiteconstraints} are satisfied. Now let $N_{ab}(L)$ be the number of inert states in a system of size $L$, in which the final two sites have $S^z$ eigenvalues $a$ and $b$ respectively. The total number of inert states for a system of size $L$ is obtained by summing $N_{ab}(L)$ over all choices $ab$. Using Table \ref{threesiteconstraints} , we can see that these quantities obey the recursion relations

\begin{eqnarray}
\left( \begin{array}{c} N_{++}\\ N_{+0}\\ N_{+-}\\ N_{0+}\\N_{00}\\N_{0-} \\N_{-+} \\N_{-0}\\N_{--}
\end{array}
\right)_{L+1} = \left(\begin{array}{ccccccccc} 1 & 0 & 0& 1 & 0 & 0 & 1 & 0 & 0 \\ 1 & 0 & 0 & 0 & 0 & 0 & 0 & 0 & 0 \\ 1 & 0 & 0 & 0 & 0 & 0 & 0 & 0 & 0 \\ 0 & 1 & 0 & 0 & 1 & 0 & 0 & 1 & 0 \\ 0 & 1 & 0 & 0 & 1 & 0 & 0 & 1 & 0 \\ 0 & 1 & 0 & 0 & 1 & 0 & 0 & 1 & 0\\ 0 & 0 & 0 & 0 & 0 & 0 & 0 & 0 & 1\\ 0 & 0 & 0 & 0 & 0 & 0 & 0 & 0 & 1\\ 0 & 0 & 1 & 0 & 0 & 1 & 0 & 0 & 1\end{array} \right)
 \left( \begin{array}{c} N_{++}\\ N_{+0}\\ N_{+-}\\ N_{0+}\\N_{00}\\N_{0-} \\N_{-+} \\N_{-0}\\N_{--}
\end{array}
\right)_{L} \nonumber
\end{eqnarray}
This matrix can be diagonalized and its eigenvalues and eigenvectors, combined with the known values for $N_{ab}(3)$ can be used to exactly determine the number of inert states for any $L$. However, asymptotically at large $L$, the growth will be controlled by the largest eigenvalue of this matrix, $\lambda$, i.e. the dimension of the Hilbert space grows asymptotically as $|\lambda|^{L}$. The matrix in question has only one real, positive eigenvalue with norm greater than one, $\lambda \approx \ 2.2$ which tells us that the dimension of the localized subspace grows asymptotically as $\sim 2.2^L$. 
\begin{table}
\begin{tabular}{c|c}
\underline{Last two sites of L site chain are} & \underline{Site added can be} \\
$+$ $+$ & $+$ or $0$ or $-$\\
$+$ $0$ & $+$ or $0$ or $-$\\
$+$ $-$ & $-$ \\
$0$ $+$ & $+$\\
$0$ $0$ & $+$ or $0$ or $-$\\
$0$ $-$ & $-$\\
$-$ $+$ & $+$ \\
$-$ $0$ & $+$ or $0$ or $-$\\
$-$ $-$& $+$ or $0$ or $-$
\end{tabular}
\caption{For the fractonic circuit with three site gates, if an inert state in a system of size $L$ has the final two sites in the states shown in the left column, then it remains inert upon addition of another spin if the new spin is in the corresponding state shown in the right column. \label{threesiteconstraints}}
\end{table}

\begin{figure*}
\centering
\includegraphics[width=\textwidth]{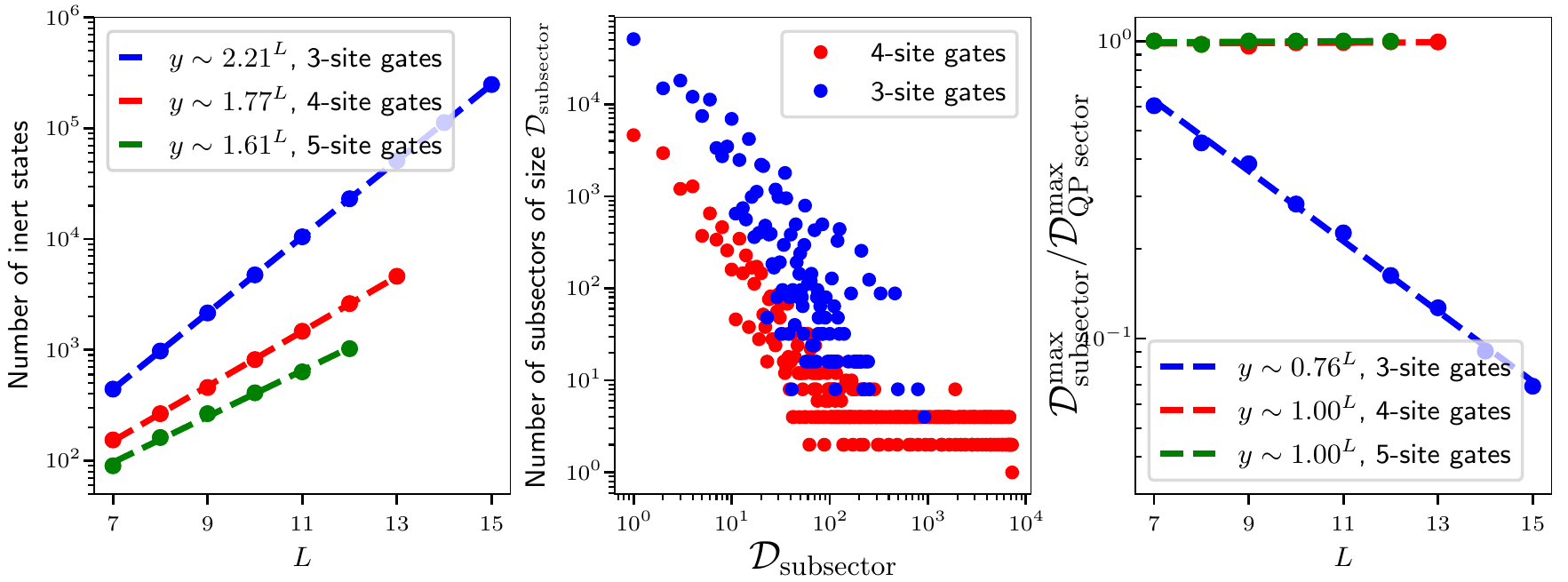}
\caption{(a) Scaling of dimension of inert subspace as a function of system size, for $N$ site gates $N=3,4,5$. For $N=3,4$ the results are consistent with the analytic predictions. For $N=5$ we have not worked out an analytic prediction, but the results are consistent with the lower bound established in Sec.~\ref{inert}.  (b) Plot showing subsector size distribution. For three site gates the frequency of subsectors of a particular size decreases polynomially with the Hilbert space dimension of the subsector. For four site gates there is an initial polynomial decrease followed by a saturation, in that beyond a certain subsector size, further increases in subsector size do not seem to translate into a decrease in frequency. Notably, the maximum size of the emergent subsectors for four site gates is much larger than that for three site gates. (c) Plot showing the relative sizes of the largest emergent subsector and largest $(Q,P)$ symmetry sector, which corresponds to $Q=0, P=0$.   For three site gates, the size of the largest emergent subsector is a vanishing fraction of the size of the largest symmetry sector, in the thermodynamic limit, whereas these sizes scale similarly for four and five site gates, showing that the fracturing is more severe for three site gates.  \label{longrangegates}}
\end{figure*}

We therefore conclude that in the thermodynamic limit there are  approximately $2.2^L$ inert states, each of which exists in its own emergent subsector, undergoes trivial (pure phase) dynamics, and does not mix with the rest of the Hilbert space. This is verified by exact numerical counting of the number of inert states in systems upto sizes $L=15$, and shown in Fig.~\ref{longrangegates}(a). 

We note that the key feature of fractonic circuits that leads to this exponentially growing inert subspace is the existence of \emph{multiple} pathways or choices for getting new inert states upon adding spins to inert states of a given size. By contrast, in a system with only charge conservation,  the only choices for building inert states require $++$ to be followed by $+$, or $--$ to be followed by $-$. This, however, gives exactly two inert states due to a lack of exponential branching arising from multiple pathways.

Generalizing this, one can verify via a similar asymptotically exact counting (see Appendix) that a fractonic circuit with four site gates also has an exponentially large localized subspace, with asymptotic dimension $\sim 1.8^L$ in the thermodynamic limit, again numerically verified in Fig.~\ref{longrangegates}(a). Exact analytical calculations for larger gate sizes rapidly become tedious, but the construction depicted in Fig.~\ref{fig:inert} is sufficient to show that an exponentially large exactly localized subspace survives for any finite gate size $N$.

\subsection{Larger subsectors}
\label{larger}
We now turn to a systematic construction of emergent dynamical subspaces of dimension greater than one, which do not mix with the rest of the Hilbert space. The main idea is to build subspaces of various sizes by embedding ``active" (non-inert) blocks into inert backgrounds, and appropriately ``shielding" the active blocks so as to keep the active region localized in a finite region of space (Fig.~\ref{fig:inert}(b)). The size of the sector so built will be controlled by the Hilbert space dimension of the active blocks, and 
we can embed multiple active blocks separated by inert regions. Strikingly, this leads to a coexistence of spatial regions that thermalize or not, starting from a single initial state! This is different even from the case of scars in other models like the PXP where the thermalization, or lack thereof, is controlled by the initial state but there is no further spatial dependence of the relaxation of observables.  

To illustrate, Fig.~\ref{fig:inert}(c) shows the expectation value of the charge $\langle S^z_x(t)\rangle$ in a system of length $L=14$, initialized in a state with a central active region surrounded by shielding regions. We can see that although the spins at the center thermalize, they never succeed in entirely melting the shielding region, so that the spins on the boundary of the system remain frozen throughout the time evolution! In other words, the shielding regions can protect the boundary spins against decoherence, despite the presence of the fluctuating active region nearby. In this example, the inert spin lies at the boundary merely for ease of depiction in a finite size system --- this chunk of 14 sites can be embedded into a larger system by extending the inert configurations on either end.

We start with some concrete examples to build intuition for how this works, and then provide a more general construction. 

A simple example for a circuit with three site gates is provided by a configuration of the form $|\cdots 0+0 \cdots \rangle$, where in each case the $\cdots $ denote inert configurations (such as the ones constructed in the previous subsection) ending with a $++$ next to the non-trivial block. 
Applying the allowed $(Q,P)$ conserving moves (Fig.~\ref{fig:floquetCircuit}) readily shows that such a state has non-trivial dynamics only over three sites in real space, and has Hilbert space dimension two. The total charge within this restricted region of real space is then independently a local integral of the motion, even though the circuit is {\it in principle} allowed to spatially move charge. Importantly, this local integral of motion is {\it state dependent} - a single charge immersed in a sea of zeros can move freely by emitting dipoles, whereas a charge blockaded on both sides by inert configurations ending in $++$ cannot leave a restricted region of real space. Multiple analogous ``active" blocks with locally non-trivial dynamics may trivially be introduced into an otherwise `inert' background, each block ``shielded" by $++$ on either end. The size of the active blocks may also be varied in size. Such constructions manifestly exist for any finite gate size, since there is always a localized subspace into which finite non-trivial blocks may be embedded, with appropriate shielding (\emph{cf.} Fig.~\ref{fig:inert}). For example, for a circuit with four site gates, $+++$ would suffice to `shield' a $0+0$ region. These are not the only examples (e.g. all charges could be reversed), but they suffice to make the point that non-trivial blocks can always be embedded into otherwise inert regions. 

\begin{figure}
\centering
\includegraphics[width=\columnwidth]{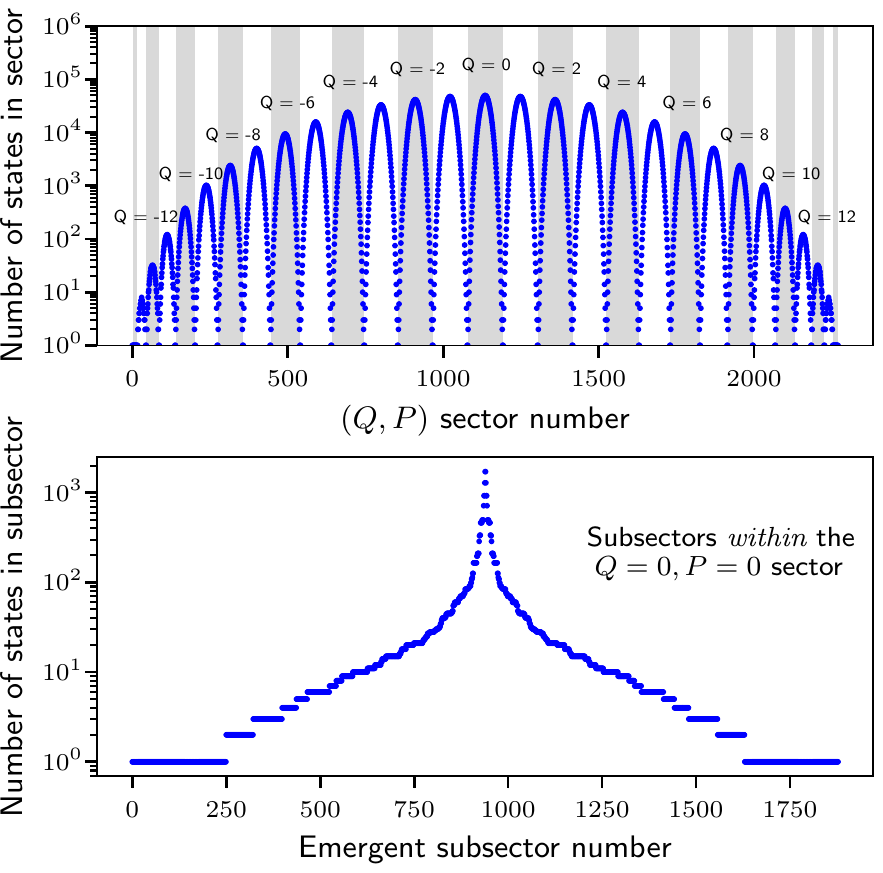}
\caption{\label{fig:subsectors} (a) Breakup of the Hilbert space into symmetry sectors labelled by charge $Q$ and dipole moment $P$ (b) Further shattering of each symmetry sector into emergent subsectors of various size, here shown for the symmetry sector with $(Q,P) = (0,0)$. }
\end{figure}

At this point one may wonder if the `embedding' of active regions into inert subspaces actually works for `active' regions of arbitrary size, or if there is a critical size of active region beyond which the problem `avalanches' \cite{avalanche}, causing the entire inert region to `melt.' However, it is straightforward to prove that {\it any} finite size of active region can always be contained by suitably chosen finite sized shielding regions. For example, take any finite sized active region, and flank it with `shielding' regions that are `all plus' to the right, and `all minus' to the left, and which are at least as large as the active region ((\emph{cf.} Fig.~\ref{fig:inert})). Now the active region can start to `melt' the shielding regions, but in doing so it will inevitably either be moving plus charge left, or minus charge right, both of which reduce the dipole moment. To preserve dipole moment overall, the active region would have to increase its internal dipole moment to compensate. However, a {\it finite} sized active region has a maximum internal dipole moment that it can accommodate, and as such the active region {\it cannot} entirely melt suitably chosen `shielding' regions of the same size. Outside the shielding regions, the state can then remain inert, as in Fig.~\ref{fig:inert}(c). At a technical level, the problem avoids avalanches \cite{avalanche} because as the `active' region grows, it has to increase its dipole moment and become less active. Consequently, one may embed active regions of any desired size into the inert subspace, by choosing the appropriate shielding. 

We have therefore proven that the Hilbert space within each symmetry sector `shatters' into numerous emergent subsectors of all sizes. This `shattering' may be straightforwardly verified numerically extracting the `connectivity' of the Floquet operator, within a particular symmetry sector. In Fig.~\ref{fig:subsectors} we show this shattering quantitatively, for a twelve site system in the sector with $Q=0$ and $P=0$ and three-site gates. The sectors with exactly one state correspond to the `inert' states (localized subspace) discussed above, but as one can see, there is a distribution of emergent subspaces of a wide variety of sizes. In Figure~\ref{longrangegates}(b), we show the full distribution of emergent subsector sizes for circuits with gate size $N = 3,4$ in a system of size $L=13$. We see that, for three site gates, the frequency of subsectors of a particular size decreases polynomially with the dimension of the subsector. For four site gates there is an initial polynomial decrease followed by a saturation, in that beyond a certain subsector size, further increases in subsector size do not seem to translate into a decrease in frequency. Notably, the maximum size of the emergent subsectors for four site gates is much larger than that for three site gates. 

In fact, the largest subsector for three-site gates is numerically observed to contain exactly $\binom{(L-1)}{\floor{(L-1)/2}}$ states, which asymptotically scales as $2^{L}$. This is a vanishing fraction of the largest $(Q,P)$ sector, which scales as $3^L$ (upto polynomial in $L$ corrections). This indicates a strongly constrained dynamics, which is only ever able to connect a vanishing fraction of the full Hilbert space, also shown quantitatively in Figure~\ref{longrangegates}(c). By contrast, the figure shows that the largest subsector with longer range gates has the same size as the largest symmetry sector, and thus the dynamics can access much larger parts of the Hilbert space. We next turn to an important feature of three site gates which may be responsible for this distinction.

\subsection{Bottlenecks and shielding}
\label{bottlenecks}
We now comment on a special feature of the fractonic circuit with three site gates, namely the existence of local integrals of motion that can act as `bottlenecks'  \emph{regardless of what larger state they are embedded into}. 

A simple example of such a bottleneck is provided by a local pattern of the form $++++$ (or the charged reversed version).  If such a (finite size) pattern is embedded into a larger state that is non-trivial everywhere to the left and the right, then the outer two $+$ charges can move away (by absorbing dipoles), but crucially these `outer' charges perfectly screen the inner charges from dipoles that could make them move. The inner $+$ charges will always be adjacent either to another $+$ charge, or to a $0$, and thus any three-site gate acting on or across the two inner charges must necessarily be trivial (pure phase). As a result, the inner two charges are perfectly localized {\it regardless} of what larger state they are embedded into, and act as a `bottleneck' that cuts the chain in two. The two halves can then be separately labeled by values of charge and dipole moment that are conserved in each half. Likewise, the presence of these bottlenecks at multiple locations can break up the chain into effectively much smaller segments, and the charge and dipole moment of each segment is separately conserved. 

On the other hand, with longer range gates, there is no finite sized motif that can `cut' the chain if embedded into an infinitely large active region. This is easiest to see if one simply embeds the finite sized motif into a sea of zeros. Suppose the left-most site of the `inert/shielding' motif is $+$ (the argument proceeds analogously if it were $-$). One may then create $-++-$ quadrupoles out of the sea of zeros, move off the $-+$ dipole to spatial infinity, and shoot the $+-$ dipole at our motif, causing the leftmost charge to move left one unit. By iterating this process, one can move the leftmost charge of the motif away to spatial infinity, leaving us with a motif reduced in size by one unit, immersed in an infinite sea of zeros (with various charges accumulated at spatial infinity). One may then repeat the process and thus `peel away' the motif one charge at a time. Accordingly, there is no finite sized motif that can `cut' the chain if embedded into an infinite active region, with longer range gates. Of course, a {\it finite} sized active region can always be contained by suitably chosen shielding regions (as discussed in Sec.\ref{larger}) and so even with longer range gates we can have patterns of finite sized `active' and `inert' regions, separated by suitably chosen shielding regions. 



\begin{figure}
\centering
\includegraphics[width=\columnwidth]{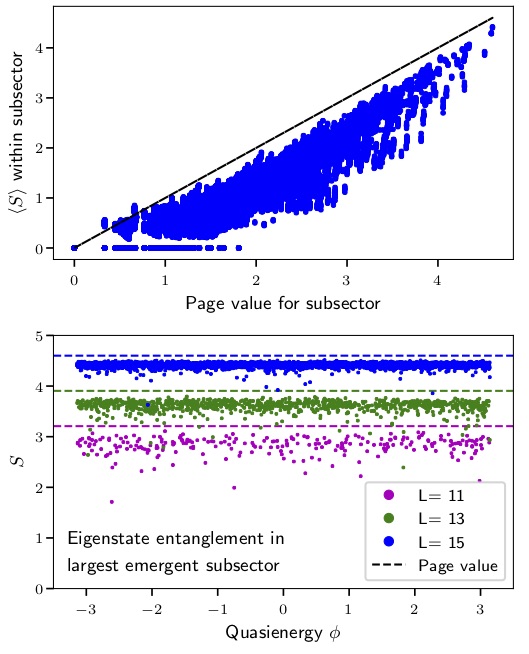}
\caption{\label{subsectors} (a) A plot showing the average bipartite entanglement entropy of eigenstates as a function of subsector size. The `Page' value would be the `thermal' entanglement entropy for a subsector of this size. Data is for $L=15$, three site gates, and open boundary conditions, and all eigenstates with $Q=0,1$ are considered. Note that while the eigenstate entanglement broadly tracks the appropriate Page value for the subsector, there is still a wide distribution, with many eigenstates having significantly subthermal half chain entanglement,  including eigenstates with strictly zero entanglement (`perfect scars') in subsectors that do not exhibit trivial dynamics. We believe this is related to the physics of bottlenecks described in Sec.\ref{bottlenecks}. (b) Entanglement entropy of individual eigenstates within the {\it largest} emergent subsector, plotted as a function of Floquet quasienergy $\phi$. Now the eigenstates do have entanglement close to the `thermal' (Page) value, and this agreement gets better as system size is increased. \label{EE2}}
\end{figure}

\subsection{Entanglement of Eigenstates}
We now re-examine our results on the mid-cut entanglement entropy of eigenstates, armed with our understanding of Hilbert space fracture. The first point is that there are emergent dynamical subsectors of varying sizes even within a single $(Q,P)$ sector, and the ``thermal" value for eigenstates in a given dynamical subsector will be controlled by the size of the subsector \cite{DonPage}. For instance, in the extreme case of strictly inert states, the eigenstate entanglement will be exactly zero. More generally, the subsectors of various sizes naturally lead to a broad distribution of high and low entanglement states -- ranging from area to volume law --  within a single extensive symmetry sector, as was observed in Fig.~\ref{EE}. 

To examine this more quantitatively, in Fig~\ref{EE2}(a), we plot the average entanglement entropy of each emergent dynamical subsector against the thermal (Page) value for that subsector in a system of length $L=15$ with three-site gates. We consider all eigenstates in all subsectors in the $Q=\{0,1\}$ sectors (with all possible $P$ values). The data is averaged over $~100$ independent circuit realizations. 
The Page value is computed by explicitly examining the $S^z$ basis states that span a given subsector, and using these to extract $\mathcal{D}_L$ and $\mathcal{D}_R$, the dimension of the Hilbert spaces in the left and right halves of the chain for that subsector. Because of the constraints, these depend on the exact basis states that form the subspace and could be different for different subsectors of the same size. Because some of the subsector sizes are very small, we use the exact expression for the Page value~\cite{DonPage} $S_{\rm Page} = \sum_{k={n+1}}^{mn} \frac{1}{k} - \frac{m-1}{2n}$, where $m = \rm{min}[\mathcal{D}_L, \mathcal{D}_R]$ and $n = \rm{max}[\mathcal{D}_L, \mathcal{D}_R]$; this reduces to the more familiar form $S_{\rm Page}\sim \log(n) - \frac{m}{2n}$ for $1 \ll m\leq n$. 

A priori one might have thought that the existence of these multiple subsectors with a broad distribution of sizes would be sufficient to explain the co-existence of high and low entanglement states within a symmetry sector. Indeed, the eigenstate entanglement does broadly track the Page value for the appropriate subsector, as shown in Fig~\ref{EE2}(a). However, the figure also shows the existence of a broad distribution of entanglement entropies {\it even after} resolving by subsector size. There even exist states with strictly zero entanglement in subsectors with dimension greater than one. Thus, the `shattering' of Hilbert space is part of the explanation for the broad distribution of entanglement entropies, but it is not the whole picture.  

This brings us to our second point --  a key part of the explanation for the broad distribution of entanglement entropies, even after resolving by subsector size, is the bottleneck/shielding phenomenon discussed in Secs.~\ref{bottlenecks}, \ref{larger}. In particular, the states with zero entanglement entropy (which are not in the strictly localized subspace) have been explicitly verified to contain a `bottleneck' motif at the midpoint of the chain, which prevents development of any entanglement across this motif, which happens to overlap the entanglement cut. The existence of such `bottleneck' motifs at positions away from the entanglement cut is also at least partially responsible for the existence of a broad distribution of entanglement entropies, even after resolving by subsector size, since the effective number of entangling degrees of freedom get reduced when the chain is `cut'. More generally, the entanglement entropy is bounded by the Hilbert space dimension of the active region that straddles the cut, and this can be much less than the size of the subsector in which the state lives, if the state consists of disconnected active regions. This discussion highlights that not only is there strong state-to-state variation in the entanglement properties of eigenstates, there is also a strong variation across spatial locations of the entanglement within a given state.

Finally, we note that the entanglement entropy in the subsector of largest size \emph{does} appear to well approximate the thermal Page value, and this agreement gets better with increasing system size (Fig.~\ref{EE2}(b)). 

\subsection{Implications for dynamics}

Finally, we turn to the implications of our results for dynamics starting from different initial states. 
While we have proven the existence of an exponentially large localized subspace, this subspace is still a measure zero fraction of the entire Hilbert space in the thermodynamic limit. While initial conditions that have high overlap with this localized subspace will clearly exhibit localization, initial conditions chosen {\it randomly} in Hilbert space will have vanishing overlap with the localized subspace. We now discuss the implications of Hilbert space shattering for the dynamics from random initial conditions. 

Dynamics from random initial conditions is expected to be highly sensitive to the degree of shattering. In Fig.\ref{longrangegates}(c) we examine what fraction of the states in a symmetry sector are contained in the emergent subsector of largest size. For three site gates, the largest emergent subsector is observed to contain a vanishing fraction of the states in the thermodynamic limit, consistent with our analytic estimates. (Recall that the largest subsector contained $\sim 2^L$ states, whereas the Hilbert space dimension is $3^L$). In contrast, for longer range gates a non-zero fraction (almost exactly equal to one) of the Hilbert space is contained in the emergent subsector of largest size, and this does not change with changing system size. 

\begin{figure}
\centering
\includegraphics[width=\columnwidth]{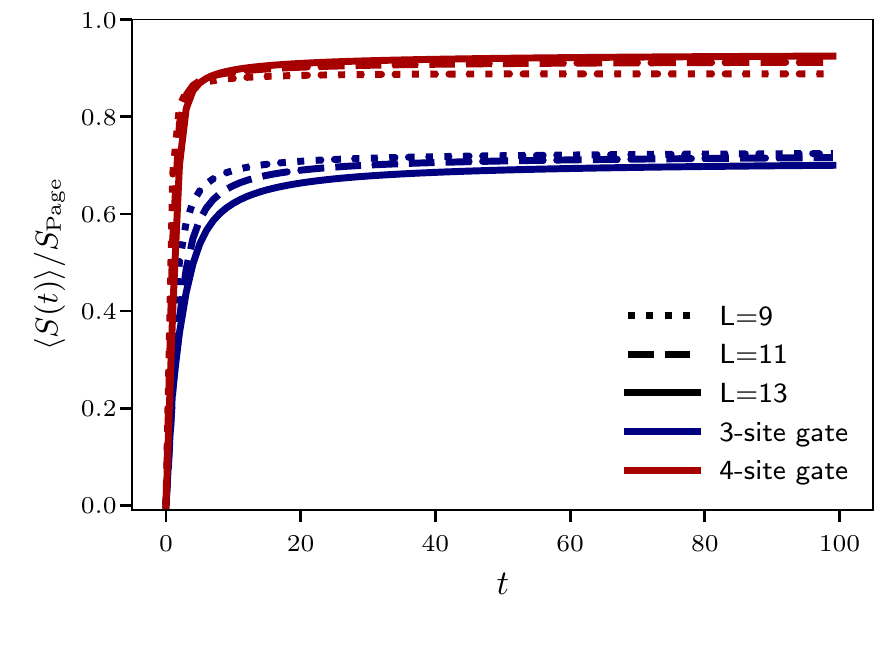}
\caption{Figure showing the dynamics of entanglement starting from a random product state (not in the $z$ basis). Entanglement entropy is given as a ratio of the `Page' value for a random product state in a Hilbert space of dimension $3^L$. For three site gates, the entanglement entropy saturates to well below its Page value, consistent with expectations given the strong fracturing of Hilbert space. For four site gates, the saturating value of entanglement entropy is much closer to the Page value, with a slow upward drift with increasing system size. \label{EEDynamics}}
\end{figure}

These differences may have interesting implications for the dynamics from randomly chosen initial product states (which are not in the $z$ basis and are not confined to any particular symmetry sector).  
For example, for three site gates, the largest subsector of Hilbert space has dimension $\sim 2^L$. The late-time entanglement entropy should therefore be dominated by this subsector and scale as $~L \ln 2$. Meanwhile, the entire Hilbert space has dimension $3^L$, and so the thermal or `Page' entanglement entropy for the full Hilbert space is of order $L \ln 3$. We would therefore expect that for a circuit with three site gates, a random initial condition should exhibit entanglement entropy growth saturating to a value approximately equal to $\frac{\ln 2}{\ln 3} S_{Page} \approx 0.63 S_{Page}.$ However, for a circuit with four site gates, the largest subsector size and the Hilbert space dimension scale similarly, as $3^L$, and one might expect dynamics starting from random initial conditions to lead to entanglement entropy growth saturating close to the Page value. (Note however a potential loophole on this argument - if the states in the largest subsector were made up of disconnected active subregions, then the saturating entropy would be bounded by the Hilbert space dimension of the active subregion straddling the entanglement cut, which could well be less than the Hilbert space dimension of the entire subsector). 

To test this intuition, in Fig.~\ref{EEDynamics}, we show the growth of entanglement entropy for both three and four site gates, starting from an initial condition that is a random product state. Note that a random product state (not in the z basis) is a superposition of multiple  symmetry sectors and subsectors. For three site gates, the entanglement entropy is observed to saturate to a clearly subthermal value of order $0.6 S_{Page}$, consistent with our expectations. Meanwhile, for four site gates the saturation value for the entanglement entropy is clearly higher, much closer to the Page value, with a slow upward drift with increasing system size. Whether the saturating value of entanglement entropy actually reaches the Page value in the thermodynamic limit is not clear from the present numerics. A more extensive investigation of pure state dynamics starting from random initial conditions, and how this depends on gate range, would be an interesting problem for future work. 

\section{Shattering (and its absence) in non-fractonic circuits}
We now point out that not all local constraints shatter Hilbert space in the manner discussed above. A good and topical example of a constrained system without a shattered Hilbert space is provided by the Rydberg chain. The dynamics of the Rydberg chain~\cite{Misha51atom} is believed to be well approximated by the PXP model \cite{turner2018weak, FendleyRydbergPhase,SachdevGirvinMott}, which acts on a constrained Hilbert space of a chain of spin $S=1/2$ variables (which may be in state $0$ or $1$),with no two adjacent spins up. That is, the Hilbert space for three sites consists of the states (0,0,0), (0,0,1), (1,0,1), (1,0,0) and (0,1,0) only. Moreover, this model only mixes the states (0,0,0) and (0,1,0), so that for a system with $L=3$ there are exactly three inert states, $(1,0,0)$, $(1,0,1)$, and $(0,0,1)$. Meanwhile if $N_{ab}(L)$ is the number of inert states in a system of size $L$ ending in $(a,b)$, then we obtain the recursion relation 
\begin{eqnarray}
\left( \begin{array}{c} N_{10} \\ N_{01} \\ N_{00}
\end{array}
\right)_{L+1} = \left(\begin{array}{ccc} 0&0&0\\ 1 & 0 & 1 \\ 1&0&0 \end{array} \right)
 \left( \begin{array}{c} N_{10} \\ N_{01} \\ N_{00}
\end{array}
\right)_{L} \nonumber
\end{eqnarray}
All eigenvalues of this matrix are zero. In sharp contrast to the fractonic circuit, therefore, the Rydberg chain does {\it not} exhibit an obvious exponentially large localized subspace or an attendant shattered Hilbert space within the constrained subspace. Nevertheless, this system does have ``scars" in this subspace, whose origin is not yet understood. These could still be caused by emergent Hilbert space fracture in some non-obvious basis, either due to emergent subsystem symmetries~\cite{ChoiAbanin} or proximity to integrability~\cite{ck} or some such mechanism. 

We note here that while the PXP model does not have any inert states in the subspace with no two adjacent spins up, it does have several inert states (and fracturing) in the full $2^L$ dimensional Hilbert space due to the limited action of the terms in the PXP model. However, as discussed previously, there is no unique or principled way to ``extend" or perturb the PXP model so this is simply a consequence of working with a special fine-tuned model (more generally, one can always write down fine-tuned Hamiltonians that leave certain subspaces invariant, but these are not expected to be stable to perturbations). By contrast the conservation or charge and dipole moment provides a physically transparent reason for the fracturing, and enables the consideration of all models respecting these constraints.

Thus far, our discussion of circuits exhibiting shattering has been particular to circuits with `fractonic' constraints (viz. conservation of charge and dipole moment).  However, not obviously fractonic circuits displaying a similar shattering of Hilbert space may also be constructed. For example, consider a circuit made out of local two spin gates acting on a one dimensional chain of $S=1$ spins. If this two site gate is constrained so that it acts trivially on the states $|0+\rangle$, $|+0\rangle$, $|0-\rangle$, and $|-0\rangle$, then it may be readily verified, through methods similar to those employed for the fractonic circuit, that there is an exponentially  large space of inert states displaying trivially localized dynamics. For a chain of size $L=2$, there are then exactly four inert states. Meanwhile, if $N_{\beta}(L)$ is the number of inert states ending in $\beta$ in a system of size $L$, then this quantity obeys the recursion relation
\begin{eqnarray}
\left( \begin{array}{c} N_+ \\ N_0 \\ N_-
\end{array}
\right)_{L+1} = \left(\begin{array}{ccc} 0 & 1 & 0 \\ 1 & 0 & 1 \\ 0 & 1 & 0 \end{array} \right)
 \left( \begin{array}{c} N_+ \\ N_0 \\ N_-
\end{array}
\right)_{L} \nonumber 
\end{eqnarray}
The matrix in the recursion relation has eigenvalues $\pm \sqrt{2}$ and zero. The dimension of the degenerate subspace thus grows asymptotically as $\sqrt{2}^L$, providing a concrete example of a not obviously fractonic circuit with an exponentially large localized subspace. The mechanism again involves the existence of ``multiple" pathways for extending inert states when new sites are added.  However, in the absence of a physical principle giving rise to this particular circuit architecture, analogous to the `fractonic' constraints of charge and dipole moment conservation, it is unclear how this circuit should be generalized to gates of longer range, and hence the question of whether this `shattering' survives in the presence of longer range gates is ill posed. Nevetheless, `shattering' may be produced by similar constructions in circuits involving gates of larger size - a sufficient condition is that there should exist at least two locally inert patterns which can be combined together in an inert fashion. 

A fruitful perspective on which types of circuits produce `shattering' of Hilbert space is provided by recursion relations of the form discussed above. For a circuit acting on a system with local Hilbert space dimension $q$, and random $N$ site gates, the recursion relation is governed by a square matrix of size $q^{N-1}$. The entries in this matrix can only be $0$ or $1$ - and at least two of the entries must be zeros, otherwise the circuit acts trivially on every possible state (which is a trivial shattering, say by diagonal matrices). {\it Every} such matrix with an eigenvalue larger than $1$ specifies a circuit with an exponentially large inert subspace. From this it follows that there are no spin $1/2$ chains with only two site gates that realize a shattered Hilbert space (in the obvious $z$ basis)- spin $S=1$ and two site gates is the minimal case necessarily to realize such shattering.

\section{Discussion}

We have shown how a local `fractonic' constraint can `shatter' Hilbert space into a huge number of emergent dynamical subsectors, leading to the emergence of exponentially large localized subspaces in which the localization is robust to temporal noise, does not require disorder, and is characterized by state dependent emergent local integrals of motion. The shattering leads to the co-existence, within a particular symmetry sector, of both high and low entanglement states similar to systems with many-body scars, thereby violating ETH as conventionally defined. Moreover the gates in the circuit may be chosen {\it randomly} subject to the constraints, so the model is not at all fine tuned. This large localized subspace could have an obvious application as a protected quantum memory. The key results have been shown to be robust for fractonic (i.e. charge and dipole conserving) circuits with {\it any} finite gate size. 

We note that insofar as the time evolution operator within each symmetry sector further `block diagonalizes' into subsectors, the analysis we have presented is reminiscent of the construction for scars by Shiraishi and Mori \cite{ShiraishiMori}. However, whereas the projective structure in \cite{ShiraishiMori} is introduced by hand, here it emerges naturally as a result of imposing `fractonic' constraints (viz. conservation of charge and dipole moment respectively). 

We have pointed out that the Rydberg chain, perhaps the best studied model in the context of `quantum scars,' does {\it not} have an (obviously) fractured Hilbert space and hence not all constraints lead to fracture.  However, we have also provided examples of not obviously fractonic circuits that exhibit shattering. What physical principles underlie these circuits - beyond the fractonic constraints discussed herein - would be an interesting topic for future work. We note that our general construction of circuits exhibiting shattering bears a striking resemblance to cellular automata, a connection that may be worth deeper exploration. We also note that a recent work exploring quantum dynamics of cellular automata demonstrated how one may construct exponentially many eigenstates in which at least some sites display trivial dynamics \cite{SarangBahti}. 

{The localization produced by Hilbert space shattering may evidently be used for information storage in the same manner as conventional MBL. Note that while storage of classical information ($S^z$) is trivial, storage of quantum information ($S^{x,y}$) is more subtle. In particular, because superpositions of `inert' product states will undergo dephasing of $S^{x,y}$, a spin echo protocol will be necessary to recover phase information, just as with MBL \cite{mblarcmp}. As with conventional MBL, however, the absence of chaos makes such a spin echo protocol in principle possible, opening up a new route to quantum information storage which (unlike MBL) does not require disorder, and is robust to temporal noise.  }

Thus far we have focused on {\it circuit} dynamics. However, {\it Hamiltonian} dynamics is of greater relevance for the study of physical systems. Insofar as Hamiltonian dynamics is `more constrained' than circuit dynamics, being required to conserve energy, we expect that it should be if anything `more localized.' Indeed our results on inert states and shielding carry over {\it mutatis mutandis} to Hamiltonian systems, with the `gates' of our circuit being replaced by the unitary time evolution operator. 
Such Hamiltonian extensions are discussed in \cite{Munich, Sanjay}. 

Additionally, it would be worth understanding to what extent the physics discussed herein carries over to higher dimensions, or to more physically realizable experimental systems. {In fact, the analytic construction encapsulated in Fig.~\ref{fig:inert} can be straightforwardly extended to higher dimensions by also imposing constraints on higher multipoles of charge. A straightforward consideration of `checkerboard' configurations analogous to Fig.~\ref{fig:inert} reveals a system in $d$ spatial dimensions, constrained to conserve the first $d+1$ multipole moments of charge, will exhibit a localized subspace exponential in system volume, as well as an analogous `shattering' of Hilbert space. This extension will be discussed at length, along with proposed realizations in ultracold atom experiments, in \cite{khn}. Techniques introduced in this manuscript have also been employed to demonstrate shattering in a `realistic' model for ultracold atom dynamics in \cite{mamaev}. }

{Our results also have important implications for the recent numerical observation of the phenomenon of `Stark MBL'~ \cite{refaelStark,PollmannStark} viz. the observation that the application of a sufficiently strong electric field can induce localization. Electric field couples to dipole moment, and a strong electric field in an energy conserving system induces dipole conservation. The effective Hamiltonian governing the system then (approximately) becomes a charge and dipole conserving Hamiltonian, which exhibits `shattering' for the reasons discussed in this paper. This shattering in turn provides an explanation for the phenomenon of `Stark MBL', illustrating that localization in these models is due to an entirely different mechanism than those considered in conventional discussions of MBL. This also furnishes a route to experimentally realizing the conservation of charge and dipole moment, at least approximately, by turning on strong electric fields. These matters will be discussed at length in \cite{Sanjay, khn}.

Of course, the broadest {\it physical} class of theories involving local constraints are gauge theories, and `fractonic' phases are known to be describable as gauge theories of `higher rank' \cite{sub}. It would be interesting to explore the possibility of Hilbert space shattering in gauge theories more generally, to clarify whether there are other types of gauge theories (beyond the `fractonic' ones discussed herein) which exhibit such shattering. This may also connect to recent works on ergodicity breaking in gauge theories \cite{NS, DS, konik, lerose, ichinose}.

Finally, we note that thus far our discussion has assumed that the constraints are applied as {\it hard} constraints, which cannot be violated. However, local constraints usually come from energetics, and are typically not `hard' but rather `soft' i.e. constraints {\it can} be violated, at the cost of paying a large energy penalty. What happens to the phenomenon of Hilbert space fracture when the constraints are softened? Presumably at the longest times the connectivity of the Hilbert space is restored, as is ergodicity and ETH, but there may well be some interesting intermediate time dynamics. Investigation of this issue would also be a fruitful topic for future work. 

{\it Note added:} While we were finalizing our manuscript, we also learned about related work by P. Sala, T. Rakovszky, R. Verresen, M. Knap and F. Pollmann which appeared in the same arXiv posting \cite{Munich}. The results of \cite{Munich} are in agreement with ours, where they overlap.  

\section*{Acknowledgments}
We would like to thank Anushya Chandran, Jason Iaconis, Chris Laumann, Sanjay Moudgalya, Shriya Pai, Abhinav Prem, Michael Pretko, and Sagar Vijay  for useful conversations and for collaborations on related work. We also acknowledge useful conversations with David Huse, Tibor Rakovszky, Pablo Sala, Shivaji Sondhi, Michael Knap and Frank Pollmann. We thank Sarang Gopalakrishnan and Romain Vasseur for feedback on the manuscript.  VK is supported by the Harvard Society of Fellows and the William F. Milton Fund. This material is based in part upon work supported by the Air Force Office of Scientific Research under award number FA9550-17-1-0183 (RN). Both authors also acknowledge the hospitality of the KITP, where part of this work was conducted, during a visit to the program ``Dynamics of Quantum Information''. The KITP is supported in part by the National Science Foundation under Grant No. NSF PHY-1748958.

\appendix 

\section{Localized subspace for fractonic circuit with four site gates}

\begin{table}
\begin{tabular}{c|c}
\underline{Last three sites of L site chain are} & \underline{Site added can be} \\
+++ & + or 0 or -\\
++0 & + or 0 or -\\
++- & - \\
+0+ & +\\
+0-& - \\
+- -&- \\
0++ & +\\
00+ & +\\
00- & -\\
0- -&- \\
-++ & +\\
-0+ & +\\
-0- & -\\
- -+ & +\\
- - 0 & + or 0 or -\\
- - - & + or 0 or -\\
+00 & - \\
-00 & +
\end{tabular}
\caption{For the fractonic circuit with four site gates, if an inert state in a system of size $L$ has the final three sites in the states shown in the left column, then it remains inert upon addition of another spin if the new spin is in the corresponding state shown in the right column. Note that we have only listed sixteen of the twenty seven possible configurations for the last three spins of the $L$ site chain - the remaining eleven configurations are `dead ends' i.e. there is nothing that can be added that leaves the state inert. \label{four}}
\end{table}

In this Appendix we provide an explicit calculation of the localized subspace for the fractonic circuit with four site gates. In this case the gates are matrices of rank $3^4=81$, with structure as detailed in Table I of \cite{pai2018localization}. Note however that there is a typo in the charge zero block of that table, in that configurations such as $+00-$ and $-00+$ should be inert, whereas $+-+-$ should mix freely with $+0-0$ and $0+0-$, but not with $+00-$. With this typo corrected, we note that in a chain of size $L=4$ there are twenty six trivial states. If a state is inert in an $L$ site system, then the addition of another site will leave it still inert as long as the last three sites of the $L$ site chain and the added site collectively form an inert state of the $L=4$ chain i.e. if the conditions detailed in Table \ref{four} are fulfilled. Note that of the twenty seven possible end states for a chain of length $L$, only eighteen allow the state to remain inert upon addition of another spin - the rest are `dead ends.' This is an important distinction to the circuit with three site gates where there were no dead ends. We can then write a recursion relation for the eighteen `live' configurations only, and it takes the form of the rank eighteen matrix equation given below.

\begin{widetext}
\begin{eqnarray}
\left( \begin{array}{c} N_{+++}\\ N_{++0}\\ N_{++-}\\ N_{+0+}\\N_{+0-} \\N_{+--} \\ N_{0++} \\ N_{00+} \\ N_{00-} \\ N_{0--} \\ N_{-++} \\ N_{-0+} \\ N_{-0-} \\ N_{- -+}\\N_{- -0}\\N_{- - -} \\ N_{+00} \\ N_{-00}
\end{array}
\right)_{L+1} = \left(\begin{array}{cccccccccccccccccc} 1 & 0 & 0 & 0 & 0 & 0 & 1 & 0 & 0 & 0 & 1 & 0 & 0 & 0 & 0 & 0 & 0 & 0 \\ 1 & 0 & 0 & 0 & 0 & 0 & 0 & 0 & 0 & 0 & 0 & 0 & 0 & 0 & 0 & 0 & 0 & 0 \\ 1 & 0 & 0 & 0 & 0 & 0 & 0 & 0 & 0 & 0 & 0 & 0 & 0 & 0 & 0 & 0 & 0 & 0  \\ 1 & 0 & 0 & 0 & 0 & 0 & 0 & 0 & 0 & 0 & 0 & 0 & 0 & 0 & 0 & 0 & 0 & 0 \\ 1 & 0 & 0 & 0 & 0 & 0 & 0 & 0 & 0 & 0 & 0 & 0 & 0 & 0 & 0 & 0 & 0 & 0 \\ 0 & 0 & 1 & 0 & 0 & 0 & 0 & 0 & 0 & 0 & 0 & 0 & 0 & 0 & 0 & 0 & 0 & 0 \\ 0 & 0 & 0 & 1 & 0 & 0 & 0 & 1 & 0 & 0 & 0 & 1 & 0 & 0 & 0 & 0 & 0 & 0 \\ 0 & 0 & 0 & 0 & 0 & 0 & 0 & 0 & 0 & 0 & 0 & 0 & 0 & 0 & 0 & 0 & 0 & 1\\ 0 & 0 & 0 & 0 & 0 & 0 & 0 & 0 & 0 & 0 & 0 & 0 & 0 & 0 & 0 & 0 & 1 & 0 \\ 0 & 0 & 0 & 0 & 1 & 0 & 0 & 0 & 1 & 0 & 0 & 0 & 1 & 0 & 0 & 0 & 0 & 0 \\ 0 & 0 & 0 & 0 & 0 & 0 & 0 & 0 &0 & 0 & 0 & 0 & 0 & 1 & 0 & 0 & 0 & 0 \\  0 & 0 & 0 & 0 & 0 & 0 & 0 & 0 &0 & 0 & 0 & 0 & 0 & 0 & 1 & 0 & 0 & 0 \\ 0 & 0 & 0 & 0 & 0 & 0 & 0 & 0 &0 & 0 & 0 & 0 & 0 & 0 & 1 & 0 & 0 & 0 \\ 0 & 0 & 0 & 0 & 0 & 0 & 0 & 0 &0 & 0 & 0 & 0 & 0 & 0 & 0 & 1 & 0 & 0 \\ 0 & 0 & 0 & 0 & 0 & 0 & 0 & 0 &0 & 0 & 0 & 0 & 0 & 0 & 0 & 1 & 0 & 0 \\ 0 & 0 & 0 & 0 & 0 & 1 & 0 & 0 & 0 & 1 & 0 & 0 & 0 & 0 & 0 & 1 & 0 & 0 \\ 0 & 1 & 0 & 0 & 0 & 0 & 0 & 0 & 0 & 0 & 0 & 0 & 0 & 0 & 0 & 0 & 0 & 0 \\  0 & 0 & 0 & 0 & 0 & 0 & 0 & 0 & 0 & 0 & 0 & 0 & 0 & 0 & 1 & 0 & 0 & 0  \end{array} \right)
\left( \begin{array}{c} N_{+++}\\ N_{++0}\\ N_{++-}\\ N_{+0+}\\N_{+0-} \\N_{+--} \\ N_{0++} \\ N_{00+} \\ N_{00-} \\ N_{0--} \\ N_{-++} \\ N_{-0+} \\ N_{-0-} \\ N_{- -+}\\N_{- -0}\\N_{- - -} \\ N_{+00} \\ N_{-00}
\end{array}
\right)_{L} \nonumber
\end{eqnarray}
The largest eigenvalue of the above matrix has magnitude $1.8$, leading us to conclude that the dimension of the localized subspace grows asymptotically as $1.8^L$, in agreement with Figure~\ref{longrangegates}(a) and again, faster than the lower bound of $2^{L/4} \sim 1.2^L$. Similar analyses may be carried through for any finite range of gates in the fractonic circuit, but the analysis rapidly becomes tedious and so we do not pursue it here. 
\end{widetext}

\bibliography{library}

\end{document}